\documentstyle{mn} 
 
\input{epsf} 
 
\newif\ifAMStwofonts 
 
\def\url#1{{\ttfamily\def\/{/\discretionary{}{}{}}#1}}

\def\Mpc{\mbox{Mpc}}

\def\LCDM{{\char'3CDM}}   
\def\mathnew{\mathsurround=0pt}   
\def\simov#1#2{\lower .5pt\vbox{\baselineskip0pt   
    \lineskip-.5pt\ialign{$\mathnew#1\hfil##\hfil$\crcr#2\crcr\sim\crcr}}}

\def\'#1{\ifx#1i{\accent"13\i}\else{\accent"13#1}\fi}

\def\mnras{MNRAS} 
\def\apj{ApJ} 
\def\aj{AJ} 
\def\prd{Phys. Rev.D} 
 
\ifoldfss 
  \ifCUPmtlplainloaded \else 
    \NewTextAlphabet{textbfit} {cmbxti10} {} 
    \NewTextAlphabet{textbfss} {cmssbx10} {} 
    \NewMathAlphabet{mathbfit} {cmbxti10} {} 
    \NewMathAlphabet{mathbfss} {cmssbx10} {} 
  \fi 
  \ifAMStwofonts 
    \ifCUPmtlplainloaded \else 
      \NewSymbolFont{upmath} {eurm10} 
      \NewSymbolFont{AMSa} {msam10} 
      \NewMathSymbol{\upi}     {0}{upmath}{19} 
      \NewMathSymbol{\umu}     {0}{upmath}{16} 
      \NewMathSymbol{\upartial}{0}{upmath}{40} 
      \NewMathSymbol{\leqslant}{3}{AMSa}{36} 
      \NewMathSymbol{\geqslant}{3}{AMSa}{3E}

    \fi 
  \fi 
\fi 
 
\ifnfssone 
  \newmathalphabet{\mathit} 
  \addtoversion{normal}{\mathit}{cmr}{m}{it} 
  \addtoversion{bold}{\mathit}{cmr}{bx}{it} 
  \newmathalphabet{\mathbfit} 
  \addtoversion{normal}{\mathbfit}{cmr}{bx}{it} 
  \addtoversion{bold}{\mathbfit}{cmr}{bx}{it} 
  \newmathalphabet{\mathbfss} 
  \addtoversion{normal}{\mathbfss}{cmss}{bx}{n} 
  \addtoversion{bold}{\mathbfss}{cmss}{bx}{n} 
  \ifAMStwofonts 
    \ifCUPmtlplainloaded \else 
      \UseAMStwoboldmath 
      \makeatletter 
      \new@mathgroup\upmath@group 
      \define@mathgroup\mv@normal\upmath@group{eur}{m}{n} 
      \define@mathgroup\mv@bold\upmath@group{eur}{b}{n} 
      \edef\UPM{\hexnumber\upmath@group} 
      \new@mathgroup\amsa@group 
      \define@mathgroup\mv@normal\amsa@group{msa}{m}{n} 
      \define@mathgroup\mv@bold\amsa@group{msa}{m}{n} 
      \edef\AMSa{\hexnumber\amsa@group} 
      \makeatother 
      \mathchardef\upi="0\UPM19 
      \mathchardef\umu="0\UPM16 
      \mathchardef\upartial="0\UPM40 
      \mathchardef\leqslant="3\AMSa36 
      \mathchardef\geqslant="3\AMSa3E 
    \fi 
  \fi 
\fi 
 
\ifnfsstwo 
  \DeclareMathAlphabet{\mathbfit}{OT1}{cmr}{bx}{it} 
  \SetMathAlphabet\mathbfit{bold}{OT1}{cmr}{bx}{it} 
  \DeclareMathAlphabet{\mathbfss}{OT1}{cmss}{bx}{n} 
  \SetMathAlphabet\mathbfss{bold}{OT1}{cmss}{bx}{n} 
  \ifAMStwofonts 
    \ifCUPmtlplainloaded \else 
      \DeclareSymbolFont{UPM}{U}{eur}{m}{n} 
      \SetSymbolFont{UPM}{bold}{U}{eur}{b}{n} 
      \DeclareSymbolFont{AMSa}{U}{msa}{m}{n} 
      \DeclareMathSymbol{\upi}{0}{UPM}{"19} 
      \DeclareMathSymbol{\umu}{0}{UPM}{"16} 
      \DeclareMathSymbol{\upartial}{0}{UPM}{"40} 
      \DeclareMathSymbol{\leqslant}{3}{AMSa}{"36} 
      \DeclareMathSymbol{\geqslant}{3}{AMSa}{"3E} 
    \fi 
  \fi 
\fi 
 
\ifCUPmtlplainloaded \else 
  \ifAMStwofonts \else 
    \def\upi{\pi} 
    \def\umu{\mu} 
    \def\upartial{\partial} 
  \fi 
\fi

\title{Tracing the Nature of Dark Energy with Galaxy Distribution} 
\author[Solevi et al.] 
{P. Solevi$^{1,2}$ \thanks{E-mail: solevi@mib.infn.it}, R. Mainini$^{1,2}$, 
  S.A. Bonometto$^{1,2}$, A.V. Macci\`o $^3$, A. Klypin$^{4}$ \&  
\newauthor{S. Gottl\"ober} $^5$ \\ 
$^1$ Physics Department G. Occhialini, Universit\`a degli Studi di 
Milano--Bicocca, Piazza della Scienza 3, I20126 Milano (Italy) \\ 
$^2$ I.N.F.N., Sezione di Milano (Italy) \\ 
$^3$  Institute for Theoretical Physics, University of Z\"urich, 
Winterturstrasse 190, CH--8057 Z\"urich (Switzerland) \\ 
$^4$ Astronomy Department, New Mexico State University, Box 30001, Department 
4500, Las Cruces, NM 88003-0001 \\ 
$^5$ Astrophysikalisches Institut Potsdam, An der Sternwarte 16,14482 Potsdam (Germany) 
} 
 
\smallskip 
 
\pubyear{2005} 
\begin{document} 
 
\maketitle 
 
\begin{abstract} 
Dynamical Dark Energy (DE) is a viable alternative to the cosmological   
constant. Yet, constructing tests to discriminate between $\Lambda$   
and dynamical   
DE models is difficult because the differences are not large. In   
this paper we explore tests based on the galaxy mass function, the   
void probability function (VPF), and the number of galaxy clusters.   
At high $z$ the number density of clusters shows large differences   
between DE models, but geometrical factors reduce the differences   
substantially.  We find that detecting a model dependence in the   
cluster redshift distribution is a hard challenge. We show that the   
galaxy redshift distribution is potentially a more sensitive   
characteristics. We do so by populating dark matter halos in $N-$body   
simulations with {\it galaxies} using well-tested Halo Occupation   
Distribution (HOD). We also estimate the Void Probability Function and   
find that, in samples with the same angular surface density of   
galaxies in different models, the VPF is almost model independent and   
cannot be used as a test for DE. Once again, geometry and cosmic   
evolution compensate each other.  By comparing VPF's for samples with   
fixed galaxy mass limits, we find measurable differences.   
\end{abstract}   
 
\begin{keywords} 
methods: analytical, numerical -- galaxies: clusters -- 
cosmology: theory -- dark energy  
\end{keywords} 
 
\section{Introduction}   
High redshift supernovae, anisotropies of the cosmic microwave   
background (CMB), as well as data on the large-scale galactic   
distribution (Riess et al. 1988, Perlmutter et al 1988, Tegmark et   
al. 2001, De Bernardis et al 2000, Hanany et al 2000, Halverson et al   
2001, Spergel et al 2003, Percival et al. 2002, Efstathiou et al 2002)   
indicate that $\sim 70\, \%$ of the world contents are due to a smooth   
component with largely negative pressure.  This component is dubbed   
dark energy (DE). Recently Macci\`o, Governato \& Horellou (2005)
presented further arguments in favor of DE based on the local ($\sim 5\Mpc$)
Hubble flow of galaxies.  The nature of DE is still open for
debate. Candidates range from a positive cosmological constant  
$\Lambda$ --~yielding a $\Lambda$CDM cosmology~-- to models with a  
slowly evolving self--interacting scalar field $\phi$ (dynamical DE;  
Ratra \& Peebles 1988; Wetterich 1988) to even more exotic physics of  
extra dimensions (e.g., Dvali \& Turner 2003).  
  
$\Lambda$CDM cosmologies are easy to study and fit most data.   
Unfortunately, to give a physical motivation to the value of   
$\Lambda$, we need a {\it fine--tuning} of vacuum energy at the end of   
the last phase transition. To rival the success of $\Lambda$CDM other   
DE models ought to predict observables hardly distinguishable from it,   
so that discriminatory tests on DE nature are not easy to devise.   
   
Up to now, most tests based on large scale structure dealt with the   
evolution of the cluster distribution. Using the Press-Schechter-type   
approximations (Press \& Schechter 1974, PS hereafter; Sheth \&   
Tormen, 1999, 2002 ST; Jenkins 2001) the expected dependence of the   
cluster mass function on DE nature was extensively studied (see, {\it   
e.g.}, Wang \& Steinhardt 1998, Haiman et al 2001, Majundar \& Mohr   
2003, Mainini, Macci\`o \& Bonometto 2003). The results were used to   
predict the redshift dependence of various observables, such as   
temperatures ($T$) or photon counts ($N$). In Sec.~2, we compare the   
(virial) mass functions for different DE.  An important -- and often   
overlooked -- factor is the dependence of halo concentration $c$ on   
the DE equation of state. For given virial mass the concentration   
varies with DE nature up to $80 \, \%$, as shown by simulations   
(Klypin, Macci\`o, Mainini \& Bonometto 2003; Linder \& Jenkins 2003;   
Kuhlen et al. 2005). The model dependence of $c$ is so strong that it   
can be used as a possible discriminatory test for strong lensing   
measurements (Dolag et al 2004, Macci\`o 2005).  

In this paper we shall focus on galactic $\sim 10^{12} h^{-1} M_\odot$
scales which are also interesting for testing models of DE.  In order
to make a prediction we first need to know how to generate a
distribution of galaxies, not just dark matter halos.  There are
different ways for doing this. We decided to use recent results on the
Halo Occupation Distribution (HOD): a probability to find $N$ galaxies
in a halo of mass $M$.  The HOD properties have been studied in
details (Seljak 2000; Benson 2001; Bullock, Wechsler \& Somerville
2002; Zheng et al. 2002; Berlind \& Weinberg 2002; Berlind et
al. 2003; Magliocchetti \& Porciani 2003, Yang, van den Bosch \& Mo
2003; van den Bosch, Yang \& Mo 2003).  Numerical simulations
including gas dynamics (White, Hernquist \& Springel 2001; Yoshikawa
et al., 2001; Pearce et al. 2001; Nagamine et al. 2001;Berlind et
al. 2003, Yang et al. 2004) or semi-analytical models of galaxy
formation (Kauffmann, Nusser \& Steinmetz 1997; Governato et al. 1998;
Kauffmann et al. 1999a,b; Benson et al. 2000a,b; Sheth \& Diaferio
2001; Somerville et al. 2001; Wechsler et al. 2001; Benson et
al. 2003a; Berlind et al. 2003) were used, to find a law telling us
how halos split in sub--halos hosting individual galaxies.
 
In this way we formulate predictions on galaxy mass functions and
their $z$--dependence.  In order to compare predictions with data we
then need to disentangle the evolution of the mass function from the
evolution of the $M/L$ ratio, because observations provide
luminosities of galaxies and our estimates give us halo masses. In
this respect, one of the aims of this work is to estimate how
precisely the $M/L$ evolution should be known, in order to use data on
galactic scales to test DE nature.  Potentially, $M/L$ evolution can
be predicted using galactic evolution models (see, e.g., Bressan,
Chiosi \& Fagotto 1994, Portinari, Sommer-Larsen \& Tantalo 2004, and
references therein). Such predictions can be compared with weak
lensing results or satellite dynamics (Prada et al 2003). The latter
methods will provide estimates of virial $M/L$ for samples of galaxies
at different $z$. This is exactly what we need in order to test
different models for DE. Here we find that, for some statistics, the
expected signal, {\it i.e.}, the differences between models are rather
large. So, there is a hope to detect DE effects in spite of
uncertainties in $M/L$ ratios.  Certainly, if galactic evolution
predictions and high--$z$ $M/L$ estimates are compared, one can hardly
prescind from taking into account accurately the impact of DE nature.
   
We run a series of $N-$body simulations with different equations of 
state. In these models the ratio $w = p_{de}/\rho_{de}$ of the DE 
pressure to the energy density varies with $z$ according to field 
dynamics. The models considered here are the Ratra-Peebles (RP, Ratra 
\& Peebles 1988) and the models with the supergravity (SUGRA, Brax \& 
Martin 1999, 2001; Brax, Martin \& Riazuelo, 2000). Appendix A gives a 
short summary on these models. Each model is specified by an 
additional parameter -- the energy scale $\Lambda$ of the 
self--interacting potential of the scalar field. Here we take $\Lambda 
= 10^3$GeV for both models. In RP (SUGRA), $w$ shows slow (fast) 
variations with $z$.

The void probability function is an obvious candidate for   
discriminating models. Fluctuations grow differently in the   
models. So, one may expect some differences in VPF.  We use galaxy   
distributions to estimate the void probability function (VPF) at   
different redshifts.  While measuring VPF in simulations is   
straightforward, mimicking observations is slightly more complicated   
because it requires corrections for geometrical effects and because   
the answer depends on a definition of galactic populations.  At $z=0$   
no model dependence of the VPF is expected or found. Predictions at   
higher $z$ depend on how galaxy samples are defined. In particular, we   
show that, if equal angular density samples are considered, VPF   
results are independent of the DE nature. On the contrary, if we   
select samples above fixed galactic mass $M$, a significant signal is   
found, which can be useful for testing the DE nature.   
   
The plan of the paper is as follows: In Sec.~2 we discuss how geometry   
and galactic evolution affect the redshift distributions of galaxies   
and clusters. In Sec.~3 and 4 we discuss the simulations and prescriptions   
of populating halos with galaxies.  In   
Sec.~5 and 6 results on redshift distributions and the VPF are   
given. Sec.~7 is finally devoted to discussing our results and future   
perspectives.   
   
\section{Geometrical and evolutionary factors}   
Let us consider a set of objects whose mass function   
is $n(>M,z)$. In a spatially flat geometry, their number between $z$   
and $z+ \Delta z$, in a unit solid angle, is given by   
\begin{equation}   
N(>M,z,\Delta z) = \int_{z}^{z+\Delta z} dz'\, D(z')\, r^2(z')\, n(>M,z')   
\label{eq:1}   
\end{equation}   
with $D(z) = dr/dz$. For flat models:   
\begin{equation}   
D(z) = {c \over H_o} \sqrt{ \Omega_m(z) \over \Omega_{mo} (1+z)^3 }~,~   
r(z) = \int_0^z dz' D(z')   
\label{eq:2}   
\end{equation}   
Here $\Omega_m(z)$ is the matter density parameter at the redshift $z$.   
The Friedman equation can be written in the form    
\begin{equation}   
\, ~H^2 (z)    
= {8\pi \over 3} G {\rho_{mo} (1+z)^3 \over \Omega_m(z)}    
= H_o^2 { \Omega_{mo} (1+z)^3 \over \Omega_m(z)}~.   
\label{eq:3}   
\end{equation}   
Along the past light cone, $a\, dr = -c\, dt$.   
So, by dividing the two sides by $dz=-da/a^2$, one finds that   
$a\, dr/dz = a^2 c\, dt/da$. Accordingly, $D(z) = {dr/ dz} =    
{c / H(z)}$ is derived from eq.~(\ref{eq:3}).   
   
When $w$ is a constant, a useful expression   
\begin{equation}   
D^2(z)   
= (c/ H_o) (1+z)^{-{3}} [\Omega_{mo} +   
(1-\Omega_{mo})(1+z)^{3w}]^{-1}   
\label{eq:6}   
\end{equation}   
can  be obtained, which allows one to see that the geometrical factors   
increase both when $w$ decreases and $\Omega_{mo}$ decreases in   
models with $w<0$.   
   
\begin{figure} 
\centering 
\epsfxsize=\hsize\epsffile{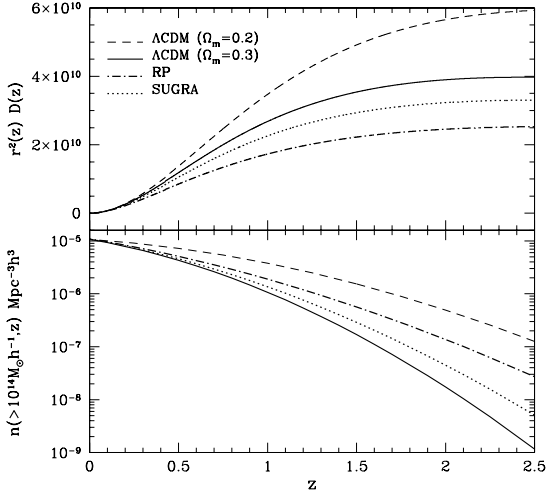} 
\caption{ 
Geometrical and evolutionary terms on the cluster mass 
scale. In both panels $\Lambda$CDM with $\Omega_{mo} = 0.2$ is above 
$\Lambda$CDM with $\Omega_{mo} = 0.3$. On the contrary, in the upper 
and lower panels RP and SUGRA lay on the opposite sides of 
$\Lambda$CDM. In the latter case, cancellation between geometrical and 
evolutionary effects is therefore expected.} 
\label{f1} 
\end{figure}

An extension to dynamical DE can be performed either by using the   
interpolating expressions yielding $\Omega_m(z)$, for RP and SUGRA   
models, provided by Mainini et al (2003b) or, equivalently, through   
direct numerical integration. We obtain the geometrical factor   
$r^2(z) D(z)$ shown in the upper panel of Fig.~\ref{f1}.   
 
\begin{figure} 
\centering 
\epsfxsize=\hsize\epsffile{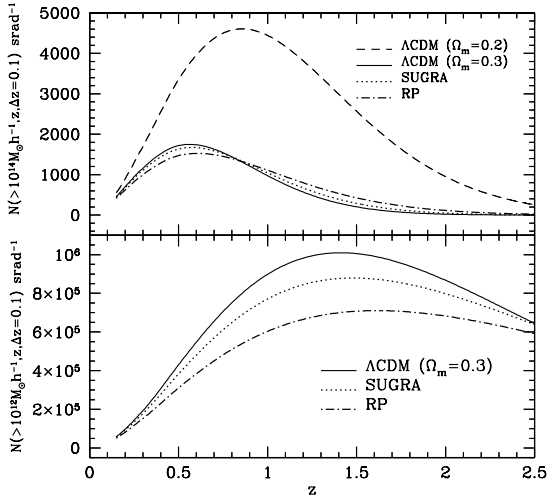} 
\caption{ 
The upper panel shows how the redshift distribution on the   
cluster mass scale depends on the models. The cancellation expected   
from Fig.~\ref{f1} has occurred and the models with $\Omega_{mo} =   
0.3$ are quite close, while $\Lambda$CDM grantes large differences   
when varying $\Omega_{mo}$. In the lower panel, similar curves are   
shown for halos on galactic mass scales. In this case, evolution is   
almost model independent and the geometrical factor causes an   
appreciable difference. On these scales a lower $\Omega_{mo}$ yields a   
different halo density, unless the spectrum is normalized to   
unphysical levels. Hence, only models with $\Omega_{mo} = 0.3$ are   
shown.} 
\label{f2}   
\end{figure}			     
Here $\Omega_{mo} = 0.3$ or 0.2 for $\Lambda$CDM, while $\Omega_{mo}   
= 0.3$ for SUGRA and RP models (for whom $\Lambda=10^3$GeV). $H_o$ is   
70$\, $km/s/Mpc in all models ($h=0.7$).  In the absence of number   
density evolution, the upper panel of Fig.~\ref{f1}  shows the   
dependence of the angular number density on $z$.   
   
In the PS formulation, the expected differential cluster number   
density $n(M)$, at a given time, is then given by the expression   
\begin{equation}   
f(\nu) \nu d\log \nu   
= {M \over \rho_m} n(M) M d\log M~.   
\label{eq:2-1}   
\end{equation}   
Here $\rho_m$ is the matter density, $\nu = \delta_c/\sigma_M$ is the   
{\it bias factor}, $M$ is the mass scale considered.  $\sigma_M$ is   
the r.m.s. density fluctuation on the scale $M$ and $\delta_c$ is the   
amplitude that, in the linear theory, fluctuations have in order that,   
assuming spherical evolution, full recollapse is attained exactly at   
the time considered (in a standard CDM model this value is $\sim$1.68;   
the difference, in other model, ranges around a few percent).  As   
usual, we took a Gaussian $f(\nu)$ distribution.   
   
Together with eq.~(\ref{eq:2-1}), we must take into account the   
virialization condition, which yields significantly different density   
contrasts $\Delta_v$ in different DE models. Further details can be   
found in Mainini et al (2003a).   
   
In the lower panel of Fig.~\ref{f1} we show the evolution of the   
number of halos of mass $>10^{14} h^{-1} M_\odot$, in comoving   
volumes. All models are normalized to the same cluster number today   
and the redshift dependence of $n(>M,z)$ is clearly understandable, on   
qualitative bases: When $\Lambda$CDM models are considered, the   
evolution is faster as we approach standard CDM. On the contrary, RP   
and SUGRA yield a slower evolution than $\Lambda$CDM.   
   
The important issue is that, while both the geometrical factor and the   
evolutionary factor of $\Lambda$CDM (with $\Omega_{mo} = 0.2$) lay above   
$\Lambda$CDM (with $\Omega_{mo} = 0.3$), RP and SUGRA factors lay on the   
opposite sides of $\Lambda$CDM.   
   
When the two factors are put together this causes the effect shown in   
the upper panel of Fig.~\ref{f2}, a strong signal on $\Omega_{mo}$   
and a widespread cancellation for DE models, compared to $\Lambda$CDM.   
Discriminating between different DE natures, from this starting point,   
is unavoidably a hard challenge.   
   
Geometrical factors do not depend on the mass scale. Instead,   
evolutionary factors are known to have a stronger dependence on the   
model for larger masses. As cancellation is almost complete on cluster   
scales, it is to be expected that geometrical factors yield a   
significant signal on lower mass scales.  This is shown on the lower   
panel of Fig.~\ref{f2}, where halos of $10^{12} h^{-1} M_\odot$ are   
considered.  A halos of this mass is expected to host a normal   
galaxy. More massive halos are expected to host many galaxies. Hence,   
this plot cannot be directly compared with observations. Its main   
significance is that such lower mass scales deserve to be inspected   
because DE signals are expected to be strong enough on these scales.   
   
\section{Simulations}   
\label{sec:sim}
The simulations run for this work are based on a $\Lambda$CDM model   
and two dynamical DE models, with the same matter density and Hubble   
parameters ($\Omega_{mo} = 0.3$ and $h = 0.7$).   
 The simulations are run   
using the Adaptive Refinement Tree code (ART; Kravtsov, Klypin \&   
Khokhlov 1997).  The ART code starts with a uniform grid, which covers   
the whole computational box. This grid defines the lowest (zeroth)   
level of resolution of the simulation.  The standard Particles-Mesh   
algorithms are used to compute density and gravitational potential on   
the zeroth-level mesh.  The ART code reaches high force resolution by   
refining all high density regions using an automated refinement   
algorithm.  The refinements are recursive: the refined regions can   
also be refined, each subsequent refinement having half of the   
previous level's cell size. This creates a hierarchy of refinement   
meshes of different resolution, size, and geometry covering regions of   
interest. Because each individual cubic cell can be refined, the shape   
of the refinement mesh can be arbitrary and match effectively the   
geometry of the region of interest.   
   
The criterion for refinement is the local density of particles: if the   
number of particles in a mesh cell (as  estimated by the Cloud-In-Cell   
method)  exceeds  the   level  $n_{\rm thresh}$,  the   cell  is split   
(``refined'')  into  8   cells of   the   next refinement  level.  The   
refinement threshold may depend on the refinement level. The code uses   
the   expansion  parameter  $a$ as  the    time variable.   During the   
integration,   spatial  refinement is    accompanied  by  temporal   
refinement.  Namely, each level of refinement, $l$, is integrated with   
its own time  step $\Delta a_l=\Delta  a_0/2^l$, where $\Delta a_0$ is   
the global time step  of the zeroth  refinement level.   This variable   
time stepping  is very important for  accuracy of the results.  As the   
force resolution  increases, more  steps  are needed to integrate  the   
trajectories accurately.  Extensive  tests of the code and comparisons   
with other  numerical $N$-body codes can  be found  in Kravtsov (1999)   
and Knebe et al. (2000).   
   
The code was modified to handle DE of different types, according to   
the prescription of Mainini et al. (2003b).  Modifications include   
effects due to the change in the rate of the expansion of the Universe   
and on initial conditions, keeping into account also spatial   
fluctuations of the scalar field before they enter the horizon.   
   
In this paper we use  4 new simulations. The models are normalized assuming   
$\sigma_8 = 0.9$. They are run in a box of 100$\, h^{-1}$Mpc. We use $256^3$   
particles with mass $m_p = 4.971 \cdot 10^9\, h^{-1} M_\odot$.  The   
nominal force resolution is 3$\, h^{-1}$kpc. All models are spatially   
flat, while $\Omega_{mo} = 0.3$ and $h=0.7$. The two $\Lambda$CDM   
models start from different random numbers and are indicated as   
$\Lambda$CDM1 and $\Lambda$CDM2. The two DE models, named RP3 and   
SUGRA3, are started from the same random numbers of $\Lambda$CDM1.   
   
\section{Galaxies in halos}   
Halos made by more than 30 particles were found in simulations by the 
same spherical overdensity (SO) algorithm used in Klypin et al 
(2003). The algorithm locates all non-overlapping largest spheres 
where the density contrast attains a given value $\Delta_v$.  Density 
contrasts are assigned the virialization values, which depend on the 
redshift $z$ and on parameters of the DE. For instance, at $z=0$, 
$\Delta_v = 101.0$ for $\Lambda$CDM, 119.4 for SUGRA3 and 140.1 for 
RP3; $\Delta_v$ values for higher $z$ can be found in Mainini, 
Macci\`o \& Bonometto (2003; see also Mainini et al 2003b). 
   
\begin{figure} 
\centering 
\epsfxsize=\hsize\epsffile{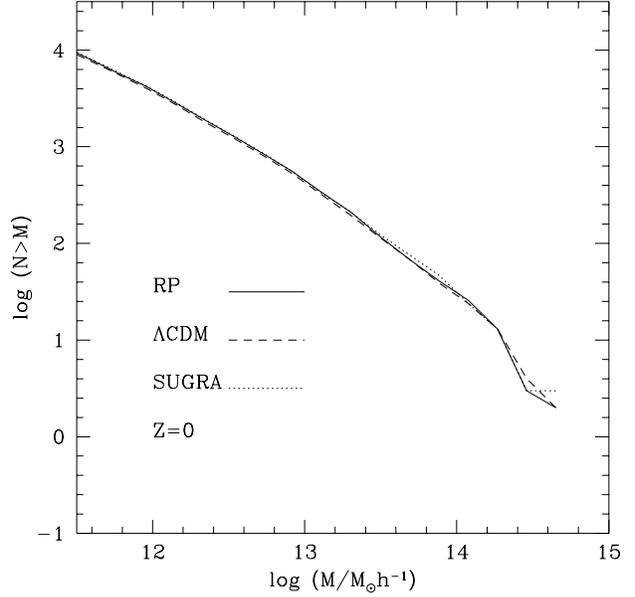} 
\caption{ 
Halo mass function at $z=0$. Results   
for \LCDM ~(dashed line), RP (solid line), and SUGRA (dotted line)   
overlap.}   
\label{f3}   
\end{figure}

Figure \ref{f3} shows the halo mass function in the LCDM1, the SUGRA3,   
and the RP3 models at $z=0$. Differences between the models are only   
due to different $w(z)$, while their $\sigma_8$ is   
identical. Accordingly, at $z=0$ their mass functions almost overlap   
and are well fitted by a the Sheth-Tormen (ST, Sheth \& Tormen, 1999)   
approximation.   
   
However, when we consider galaxies, we need to use another approach 
because individual halos may host many galaxies of different mass and 
luminosity. In order to assign galaxies to halos we use a HOD. 
This is a relatively novel approach to locate galaxies in each halo.   
It can be used in different ways. Full-scale semi-analytical methods 
can predict such quantities as the luminosity, colors, star-formation 
rates. Unfortunately, many important mechanisms are still poorly 
understood making the results less reliable.  It seems therefore 
advisable to minimize the physical input, keeping just to 
gravitational dynamics. 
   
\begin{figure} 
\centering 
\epsfxsize=\hsize\epsffile{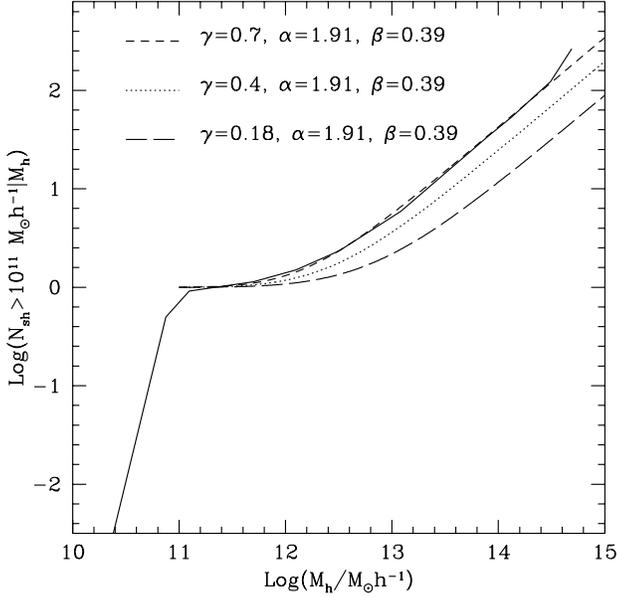} 
\caption{ 
Comparison of different halo occupation distributions.    
The  solid line is the HOD given by Kravtsov et al   
(2004). Other curves are obtained from eq.(8)  for   
different  parameters $\gamma$ as indicated in the plot.}   
\label{f4}   
\end{figure}

In this paper we use a prescription consistent with the results of 
Kravtsov et al (2004). We utilize an analytical expression recently 
proposed by Vale \& Ostriker (2004), but parameters of the 
approximation are different form Vale \& Ostriker (2004) and tuned to 
produce a good fit to results of Kravtsov et al (2004). The 
approximation is based on the assumption that the probability 
$P_s(N_s|M)$ for a halo of mass $M$ to host $N_s$ subhalos~ is 
approximately universal. We take the Schechter approximation 
\begin{equation} \label{nsubhalo}   
N(m|M)\, dm = A\, \frac{dm}{\beta M}    
\, \Big(\frac{m}{ \beta M}\Big)^{-\alpha}    
\exp\Big(-\frac{m}{ \beta M}\Big)    
\label{hod1}   
\end{equation}   
for the number of subhaloes with masses in the range $m$ to $m+dm$,   
for a parent halo of mass $M$. $A$ must be such that the total mass in   
subhaloes, $\int_{0}^{\infty}dm\, m N(m|M)$, is a fraction $\gamma M$   
of the parent halo mass. Therefore $   
A= \gamma / \beta \Gamma(2-\alpha)$, so that   
the number of sub--halos of mass $m$ is   
\begin{equation}    
\label{intsubhalomf}   
n_{sh}(m) = \int_{0}^{\infty} dM\,  N(m|M)  n_h(M)~,   
\label{nsh}   
\end{equation}   
($n_h(M)$ is the halo mass function) independently from the parent   
halo mass. The expression (\ref{hod1}) yields the following number of   
subhaloes with mass $>m$ in a halo of mass $M$:   
\begin{equation}    
\label{occupation}   
N_{sh} (>m,M)   
=\frac{\gamma}{\beta\Gamma(2-\alpha)}   
\int_{m/\beta M}^\infty dx~x^{-\alpha}    
\exp(-x)~.   
\end{equation}   
Sub--haloes will be then identified with galaxies. 
Figure \ref{f4} shows that the expression approximates the results of   
Kravtsov et al (2004) once we fit parameters $\alpha, \beta, \gamma$   
and add to the expression (\ref{occupation}) a unity, {\it i.e.} the   
halo as a sub--halo of itself.  
For large haloes, to be interpreted as galaxy clusters, this sub--halo 
could be the central cD; but adding an extra object, in such large 
galaxy sets, is just a marginal reset. For small haloes, instead, it 
is important not to forget that they represent a galaxy, as soon as 
they exceed the galaxy mass threshold. 
 
In a different context, Vale \&   
Ostriker (2004) use the value $\gamma = 0.18$, $\beta = 0.39$.   
Owing to the use we make of eq.~(\ref{occupation}), $\gamma = 0.7$   
appears more adequate.   
   
If the (differential) halo mass function $n_h(M)$ is  known, the   
sub--halo mass function is   
$$   
N_{sh}(>m) = {\gamma \over \beta \Gamma(2-\alpha)}   
\times ~~~~~~~~~~~~~~~~~~~~~~~~~~~   
$$   
\begin{equation}   
\times \int_{m/\gamma}^\infty   
dM \, n_h(M) \int_{m/\beta M}^\infty dx\, x^{-\alpha} \exp(-x)~.   
\label{Nsh}   
\end{equation}   
If we perform nonlinear predictions, $n_h(M)$ is obtained from   
the expression (\ref{eq:2-1}) or from the corresponding expressions in   
the ST case.  When we deal with simulations, halo masses have discrete   
values $m_\nu = \nu m_p$, appearing $n_h^{(\nu)}$ times, up to a top   
mass $\nu_M m_p$. Then   
$$   
N_{sh}(>m) = {\gamma \over \beta \Gamma(2-\alpha)} \times   
~~~~~~~~~~~~~~~~~~~~~~~~~~~   
$$   
\begin{equation}   
\times \sum_{\nu={m \over \gamma m_p}}^{\nu_M}   
 n_h^{(\nu)} \int_{\nu m/\beta M}^\infty dx\, x^{-\alpha} \exp(-x)~.   
~~~~~   
\label{Nshd}   
\end{equation}   
   
\begin{figure} 
\centering 
\epsfxsize=\hsize\epsffile{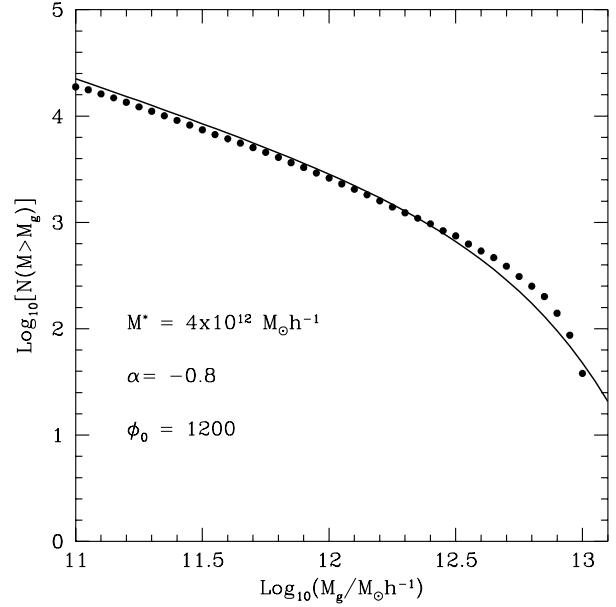} 
\caption{ 
Galaxy (cumulative) mass function at $z=0$ from simulations   
(dots) {\it vs.}~an integral Schechter function, with $\Phi_o$,   
$\alpha$ and $M_*$ shown in the frame.   
}   
\label{f5}   
\end{figure}

In Figure \ref{f5} we plot the galaxy mass function obtained with
eq.~(10), using the mass function of halos in the simulations, and
identifying sub--haloes with galaxies. At $z=0$ model differences are
unappreciable and the plotted function holds for all models.  We also
plot a Schechter function with the parameters shown in the frame,
selected to minimize the ratios between differential values at all
points.  As expected, the two curves are close. In fact, there must be
some relation between masses and luminosities, but the $M_g/L_g$ ratio
should not be a constant.
   
\section{Galaxy angular density in models with different DE}   
Let us now consider the galaxy mass function at higher $z$, for the   
different models. According to eq.~(\ref{eq:1}), the number of   
galaxies with mass $>m$, in a solid angle $\Delta \theta^2$   
($\Delta \theta \ll 1$), between redshifts $z$ and $z+\Delta z$, is   
\begin{equation}   
{N_g(>m,z;\Delta z,\Delta \theta) \over \Delta z \Delta \theta^2  }   
\simeq c\, {r^2 (z) \over H(z)} n_{g}(>m,z)~,   
\label{eq:ngal}   
\end{equation}   
if $n_{g}(>m,z)$ is the comoving number density of galaxies with mass   
$>m$ at a redshift $z$. The galaxy density can be obtained from the   
subhalo mass functions ({\ref{Nsh}) and ({\ref{Nshd}).  Accordingly,   
the average angular distance $\theta_{gg}$ is given by   
\begin{equation}   
\theta_{gg}(>m,z) \sqrt{\Delta z} \simeq {1 \over r(z)}   
\left[ H(z) \over n_{g} (>m,z) \right]^{1/2}~.   
\label{thgg}   
\end{equation}   
Thus, the l.h.s. expression is independent of the particular   
volume considered.     
   
Such $\theta_{gg}(>m,z)$ therefore depends on geometry, halo mass   
function and HOD. We however expect that the redshift dependence   
mostly arises from geometry, while evolution plays a significant role   
at higher $z$. In fact, the main difference between this and the   
cluster case is that evolution is mild and discrepancies between   
models, in comoving volumes, up to $z \sim 2$, are even weaker.   
   
Let  $\theta_{\Lambda}(z),~\theta_{SU}(z),~\theta_{RP}(z)$ be the   
mean angular distances between galaxies at redshift $z$ for the   
$\Lambda$CDM, SUGRA and RP models, respectively. Besides  these   
functions, let us also consider the angular distance $\theta_{geo}(z)$   
obtained from eq.~(\ref{thgg}), keeping the value of $n_g(>m,z=0)$ at   
any redshift and the $\Lambda$CDM geometry. Therefore,   
$\theta_{geo}(z)$, although the symbol has no reference to   
$\Lambda$CDM, describes the behavior of the angular separation {\it in   
a {\rm $\Lambda$CDM} model}, in the limit of no halo evolution.   
   
In Figure \ref{f6}, we compare results obtained from simulations with   
the ST predictions in different models.  Rather than presenting   
$\theta_{mod}(z)$, we plot the fractional difference $(\theta_{mod} -   
\theta_{geo})/\theta_{geo}$. In principle, error bars can be evaluated   
in two ways: (i) by comparing $\Lambda$CDM1 with $\Lambda$CDM2 (cosmic   
variance) and (ii) by comparing the differences between models. The   
latter can be done only at present because the models have the same   
power spectrum only at $z=0$. The differences between models exist   
because the fluctuations grow differently in the past.  At larger $z$,   
this {\it evolutionary variance} should be smaller, but is not easy to   
evaluate. We use differences between models at $z=0$ as a rough   
estimate of error bars at all redshifts. Judging by the differences   
between $\Lambda$CDM1 and $\Lambda$CDM2, the cosmic variance seems to   
be smaller by a factor of three than the evolutionary differences.  We   
find similar behaviour for different galaxy masses.  In all cases,   
differences between models can be clearly seen.   
   
The largest differences between models are attained at $z \sim   
1$. Let us remind that the plot shows the fractional differences   
between DE models and the $z$--dependence due to the mere $\Lambda$CDM   
geometry. At $z \simeq 1$, this difference is just $\simeq 5\, \%$ for   
$\Lambda$CDM while, for SUGRA, it is $\sim 20\, \%$, because of the   
different geometry and a still slower cosmological evolution. Even   
larger is the difference with RP. This compares with an evolutionary
variance hardly exceeding $\sim 2\, \%$, if the effective comoving   
volume inspected is $\sim 10^6 h^{-3}$Mpc$^3$.  For $\Delta z \sim   
0.1$, this corresponds to $\delta \theta \sim 30$--$40^o$.   
\begin{figure} 
\centering 
\epsfxsize=\hsize\epsffile{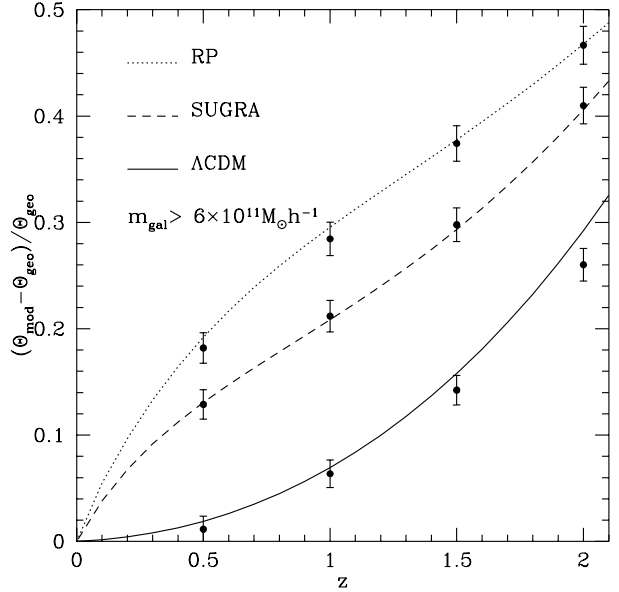} 
\caption{ 
The redshift dependence of the mean fractional angular   
separation in different models. Points with error bars show results of   
simulations. Curves indicate ST predictions.   
 The plot shows  a strong dependence   
of the galaxy redshift distribution on the DE nature.}   
\label{f6}   
\end{figure}   

The discriminating power of this theoretical prediction is to be still
compared with two possible sources of error: (i) Peculiar velocities,
setting individual galaxies into an apparent redshift band different
from the one to which they belong. (ii) Luminosity evolution.

Overcoming the latter point is critical to the use of galaxies as
indicators of DE nature and we shall devote the whole next section to
the impact of luminosity evolution. The conclusion of this discussion
is that galaxies are indeed a possible indicator of DE nature, but
more work is needed before they can be efficiently used to this aim.

In this section we shall report the result of a test performed to
evaluate the impact of redshift displacements due to peculiar
velocities. 

We divided the volume $L^3$ of the box into cells of side $10^{-1} L$,
with volume $10^{-3}L^3$. 50 such cells were selected at random at
each redshift and replaced by the cells at the closest redshift
considered with their galaxy contents. The redshift displacement
between the original cell and its replacements is $\pm 0.5$. For a
change of $5\, \%$ in galaxy contents the average shift of
$\theta_{gg}$ is $\sim 1.2\, \%$.

This shift lays well below the error bars shown in Figure \ref{f6}.
 However, the average errors due to evolutionary and cosmic
variance plus peculiar velocities are $\sim 2.4\, \%$ and the whole
error never exceeds 2.6--2.7$\, \%\, $. We therefore argue that these
sources of error do not affect the robustness of results.

\section{Dependence on mass--luminosity relation}
Let us now consider galaxy evolution, which is expected to cause
$z$--dependence of the average $M_g/L_g$ ratio, but can also yield a
$z$--dependence of the $M_g/L_g$ distribution about such average, in a
way which may depend on the mass range considered. From an
observational point of view, when we consider galaxies of various
luminosities $L_g$, we must then take into account that their expected
mass $M_g$ could be distributed with different laws at different $z$
and $L_g\, $.

\rm
Let us therefore consider the galaxy distribution on the $M_g,L_g$
plane at a given $z$, yielding the galaxy number
\begin{equation}
dN = D(M_g,L_g,[z]) \, dM_g\, dL_g
\label{nml1}
\end{equation}
in the infinitesimal area $ dM_g\, dL_g$ about the point $ M_g,L_g\,
$. We put $z$ in brackets to outline that, in respect to it, $D$
is not a {\it distribution} but a {\it function}$\, $.  Obviously we
expect a strong correlation between $ M_g$ and $L_g\, $, at any $z$,
 so that it makes sense to consider an average $M_g/L_g$ ratio.

Once the distribution $D$ is assigned, the distributions on $M_g$ ($\,
$at fixed $L_g\, $) and on $L_g$ (at fixed $M_g\, $) read
$$
\phi(M_g,[z]) = \int dL_g \, D(M_g,L_g,[z])~,
$$$$
\psi(L_g,[z]) = \int dM_g \, D(M_g,L_g,[z]).
$$
The number $dN$ is also the product of $\phi(M_g,[z])$ times the
distribution on luminosities at fixed mass $M_g$:
\begin{equation}
dN = \phi(M_g,[z])\, Q(L_g;[M_g,z])\, dM_g\, dL_g~.
\label{nml2}
\end{equation}
Equating the r.h.s.'s of eqs.~(\ref{nml1}) and (\ref{nml2}) yields
\begin{equation}
Q(L_g;[M_g,z]) = {D(M_g,L_g,[z]) \over \int dl\, D(M_g,l,[z])}
\label{qlm}
\end{equation}
and, similarly, the distribution on masses at given $L_g$, reads
\begin{equation}
\label{pml}
P(M_g;[L_g,z]) = {D(M_g,L_g,[z]) \over \int dm\, D(m,L_g,[z])}~.
\end{equation}
We can now use $P$ to work out the average $M_g/L_g$ at fixed $L_g$, and the
distribution on $M_g/L_g$ about such average. Clearly
$$
\left\langle M_g \over L_g \right\rangle_{L_g,z} =
{1 \over L_g} \int dm\, m\, P(m;[L_g,z]) = ~~~~~~~~~~~~~~~~~~~~~~~~~~~~~
$$
\begin{equation}
~~~~~~~~~~~~~~~\, = {1 \over L_g} {\int dm\, m\, D(m,L_g,[z]) \over \int
dm\, D(m,L_g,[z])}~,
\label{eq5}
\end{equation}
while the distribution
\begin{equation}
{\cal D}(M_g;[L_g,z]) = {M_g \over L_g} {D(M_g,L_g,[z]) 
\over \int dm\, D(m,L_g,[z])}
\label{eq6}
\end{equation}
tells us how $M_g/L_g$ is distributed around $\left\langle M_g/ L_g
\right\rangle_{L_g,z} \,$. 

The impact of the evolution of stellar populations (or other
mechanisms) on the $M_g/L_g$ ratio can be fully expressed through
the distribution $D(M_g,L_g;[z])$ in eq.~(\ref{nml1}).  From it
we can work out an average mass/luminosity ratio $\left\langle M_g/L_g
\right\rangle \,$ and the distribution on masses $\cal D$; they both
depend on $L_g$ and $z$.

Let us now try to inspect how such variable $\cal D$ distribution
affects our results, taking into account that we mostly ignore how
such variations occur. Accordingly, we shall proceed as follows: we
define a ``wild'' distribution, that we expect to spread the
$M_g/L_g$ ratio, at fixed $L_g$, farther from average than any
physical $\cal D$, at any $z$, will do. The effects caused by such
wild distribution should then be an overestimate of the effects of the
actual distributions.  Should they cause just a minor perturbation in
estimates, all we have to worry about is the redshift dependence
of the average $\left\langle M_g/L_g \right\rangle_{L_g} \,$.

More in detail, we shall allow that a galaxy of given mass $M_g$ has a
luminosity in an interval $L_1,L_2$ with $L_2 \sim 20\, L_1$. We test
this prescription without direct reference to luminosities: In the
sample of galaxies obtained through the HOD, at each $z$, each galaxy
mass ($m$) is replaced by a mass $m' = m + \Delta m R$, $R$ being a
random number with normal distribution and unit variance. We take
$\Delta m = 0.8\, m$, but replace all $m' < 0.1\, m$ with $0.1\, m$,
as well as all $m' > 1.9\, m$ with $1.9\, m$. The shift is therefore
symmetric on $m$ (not on $\log m$). The operation causes a slight
increase of the mass function above $\bar M \sim 2.8 \cdot 10^{11}
h^{-1} M_\odot$ (by a few percents), as there are however more lighter
galaxies coming upwards than heavier galaxies going downwards (below
$\bar M$, the low--mass cut--off of the mass function, set by the mass
resolution of our simulations --$\, $see section \ref{sec:sim}$\, $--
begins to cause shortage of transfers upwards). The operation is then
completed by reducing all masses by a (small) constant factor, so
that, summing up all masses of objects with mass $M_g > \bar M$, we
have the same total mass as before the operation. This lowers the
limit below which the mass function preserves its initial shape, but
we never use galaxy samples including masses below $ 3 \cdot 10^{11}
h^{-1} M_\odot$.

We re--estimated $\theta_{mod} \sqrt{\Delta z}$ using the new masses,
for the same mass limits as before, and compare the changes obtained in
this way with the Poisson uncertainty, due to the finite number of
galaxies in each sample.

We find that the error obtained from the above procedure ranges
between 20 and 40$\, \%$ of the Poisson error.

This output tells us that the evolution of the physical distribution
can be expected to redistribute results well inside Poisson
uncertainty. If we expect this redistribution to be random, the
top value of the whole expected error is then 3$\, \%$; if we
attribute a systematic character to it and refrain from performing a
quadratic sum with the other error sources considered, the overall
possible error is still within 3.8$\, \%$. It should be outlined that
here we pushed all error sources to their maximum; thus, we believe
that the above estimates are safely conservative.

Let us now discuss how the evolution of the average $M_g/L_g$ ratio
can affect the use of galaxies to detect DE nature. In principle, one
could use the $z$ dependence of $D(M_g,L_g)$, to obtain the
$z$--dependence of the $\langle M_g/L_g \rangle_{L_g}$ ratio. More
realistically, suitable data sets can directly provide the
$z$--dependence of $\langle M_g/L_g \rangle_{L_g}$, with some residual
uncertainty. The basic issue is then: How well the $z$ evolution of
$\langle M_g/L_g \rangle_{L_g}$ is to be known, in order that we can
test DE nature$\, $?
 
Figure \ref{f6} is however devised so to provide a direct reply to
this question: If we double the values of $\delta \theta_{gg}
/\theta_{gg}$ provided there, we have a fair estimate the difference
between evolution rates of ${\cal R} = \langle M_g/L_g
\rangle_{L_g}(z)/ \langle M_g/L_g \rangle_{L_g} (z=0)$ needed just to
cover the differences between geometry and dynamics for the
$\Lambda$CDM model or to compensate the differences between the
$\Lambda$CDM geometry and whole evolution for dynamical DE models, (in
the limit $M_g \ll M_*$, where $M_*$ is the mass scale appearing in a
Schechter--like expression). Figure \ref{f6} can just be interpreted
in both senses, changing the name of the ordinate.

For instance, at $z=0.5$, an uncertainty $\sim 20\, \%$ ($ 35\, \%$)
on the evolution of $M_g/L_g$ is needed to hide the difference between
$\Lambda$CDM and SUGRA (RP).

Let us show this point. Owing to eq.~(\ref{thgg}), $\theta_{gg}
\propto n_g^{-1/2}$, so that a shift $\delta \theta_{gg}$ in the
observed angular distance arises from a shift
\begin{equation}
{\delta n_g \over n_g} \simeq 2 {\delta \theta_{gg} \over  \theta_{gg}}
\label{thetan}
\end{equation}
in the galaxy number density 
\begin{equation}
n_g (>M_g) = n_g \left( >L_g {M_g \over L_g}\right) ~. 
\label{ngl} 
\end{equation} 
The latter shift, as shown in eq.~(\ref{ngl}), can arise from a shift 
on ${M_g / L_g}$, being 
\begin{equation} 
\delta n_g = n_g (M_g) L_g\, \delta \left(M_g \over L_g\right)~; 
\label{dng} 
\end{equation} 
here $n_g (M_g)$ is the differential mass function, obtained 
by differentiating the integral mass function $n_g (>M_g)$. 
Therefore, 
\begin{equation} 
{ \delta n_g(>M_g) \over M_g n_g(M_g) } 
\simeq  
{\delta (M_g / L_g) \over (M_g / L_g)} 
\end{equation} 
and 
\begin{equation} 
{\delta (M_g / L_g) \over (M_g / L_g)} 
\simeq 
{ \delta n_g(>M_g) \over n_g(>M_g)} 
{n_g(>M_g) \over M_g n_g(M_g) }~. 
\label{mvsn}
\end{equation} 
If we approximate the integral mass function by a Schechter 
expression, it is $|m\, n(m)/ n(>m)| = 1+{m / M_*}$, so that 
\begin{equation} 
{\delta (M_g / L_g) \over (M_g / L_g)} 
\simeq 
{ \delta n_g(>M_g) \over n_g(>M_g)} 
\left( 1+{M_g \over M_*} \right )^{-1} ~. 
\end{equation} 
Using this equation together with eqs.~(\ref{thetan}) and (\ref{mvsn}), we
have the relation
\begin{equation}
{ \delta (M_g/L_g) \over (M_g/L_g) } 
\simeq
2 \left( 1+{M_g \over M_*} \right)^{-1} 
{ \delta \theta_{gg} \over \theta_{gg} } 
\label{uncert}
\end{equation}
telling us how to use Figure \ref{f6} to estimate the evolution of
$M_g/L_g$ needed to yield the same effects of a change in DE
nature. This equation also tells us how to use Figure \ref{f6} for
masses approaching $M_*$.

\section{The Void Probability Function}     
Let us randomly throw spheres of radius $R$, in a space where objects 
of various masses $M$ are set. The probability $P_o(R)$ of finding no 
object with $M > M_{tr}$ in them, if the VPF for objects of mass 
$>M_{tr}$. 
 
We expect and find no model dependence in the {\it galaxy} VPF's at 
$z=0$. At $z > 0$, a critical issue is how $M_{tr}$ is set.  One can 
simply plan to determine the galaxy masses $M_g$ from data (e.g., from 
$L_g$ values), so to select galaxies with $M_g > M_{tr}$. As widely 
outlined, this choice involves several complications. Another option 
is taking the most {\it luminous} galaxies up to an average angular 
distance $\theta_{gg}$. 
 
Each threshold $M_{tr}$, for any $z$ and $\Delta z$, yields a value of 
$\theta_{gg}$. Fig.~\ref{f6} shows how $\theta_{gg}$ depends on the 
model at fixed threshold. {\it Viceversa}, if we keep, for that $z$ 
and $\Delta z$, a fixed $\theta_{gg}$, the relative $M_{tr}$ varies 
with models. We can compare models either at fixed $M_{tr}$ or at 
fixed $\theta_{gg}$. Dealing with observations, the latter option is 
easier, but mixes up the intrinsical VPF dependence on the model and 
other features, which also depend on the model. 
 
Besides of threshold setting, another issue bears a great operational 
relevance. In principle, VPF's can be evaluated in the comoving 
volumes where galaxies are set and compared there with VPF's from 
data. This is however unadequate to evaluate how discriminatory is the 
VPF statistics. To do so, we rather follow the following steps: 
 
\noindent   
(i) From cartesian coordinates $\bar x_i$ ($i=1,2,3$) in comoving 
volumes we work out the redshift and the celestial coordinates 
$z,~\theta,~\phi$ that an observer, set at $z=0$, would measure. This 
is done by using the geometry of the model. 
 
\noindent   
(ii) Data also give $z,~\theta,~\phi$, for each galaxy.  But, to 
estimate the VPF, they must be translated into cartesian coordinates 
$x_i$. An observer can only perform such translation by using the 
geometry of a {\it fiducial} model, {\it e.g.}, $\Lambda$CDM. The 
second step, to forge predictions, therefore amounts to re--transform 
$z,~\theta,~\phi$ into cartesian coordinates $x_i$, but using now the 
{\it fiducial} $\Lambda$CDM geometry; $x_i$'s coincide with the $\bar 
x_i$'s only for a $\Lambda$CDM cosmology. Let us call {\it fiducial} 
space, the environment where galaxies are now set. 
   
\noindent   
(iii) We then estimate the VPF's, for all models, in the fiducial 
space; these VPF's should be compared with observational data, but can 
also be compared one another to assess how discriminatory this 
statistics can be. 
 
Although comparing predictions for VPF's in comoving volumes,
therefore, bears little discriminatory meaning, our outputs are more
easily explained if we start from comoving space VPF's. Let us recall
that, on galaxy scales, evolutionary differences, up to $z \sim 2$,
are modest. Accordingly, for an assigned $M_{tr}$, we find just
marginal discrepancies, as is shown in Fig.~\ref{f6ter} for $M_{tr} =
6 \cdot 10^{11} M_\odot h^{-1}$.
Differences among VPF's arise, of course, if we do not fix $M_{tr}$, 
but $\theta_{gg}$, as a reflex of $\theta_{gg}$ differences. These 
VPF's are shown in Fig.~\ref{f7}. 
 
VPF's in respect of comoving coordinates bear a strict analogy with
mass functions in comoving volumes. The fact that geometry erases
almost any signal on the cluster mass function is analogous to what
happens for the VPF, when we pass from the comoving to the fiducial
space.  This fact is far from trivial. Let us compare these VPF's with
the VPF's in a Poisson sample with the same $\theta_{gg}$
(Fig.~\ref{f8v}) and remind that (White 1979)
\begin{equation} 
P_o(R) = \exp \left[ -\bar N_R + 
\sum_{n=2}^\infty{(-\bar N_R)^n \over n!} \xi^{(n)}(R) \right]~. 
\label{po} 
\end{equation} 
Here $\bar N_R$ is the average number of points in a sphere of radius 
$R$; $\xi^{(n)} (R)$ are the $n$--point functions averaged within the 
same sphere. For the Poisson sample $P_o(R) = \exp(-\bar N_R)$, as all 
$\xi^{(n)} (R)$ vanish. The difference between Poisson VPF and model 
VPF is to be fully ascribed to $\xi^{(n)}$, as $\bar N_R$ is set 
equal. This difference is huge, in respect to the differences between 
models, which should arise because of $\xi^{(n)}$ shifts. Their 
paucity indicates that density renormalization almost erases the 
shifts in correlation functions of all orders.

The cancellation between geometrical and $\theta_{gg}$ effects, shown 
in Fig.~\ref{f8d}, indicates that the passage from comoving to 
fiducial coordinates bears a weight comparable with the differences 
shown in Fig.~\ref{f7}. It therefore comes as no surprise that VPF's, 
for fixed $M_{tr}$, almost absent in the comoving space, are 
significant in the fiducial space. They are shown in Fig.~\ref{f9} 
and, as is expected, the curves of different models appear in 
the opposite order in respect to Fig.~\ref{f7}. 
 
If the redshift dependence of the $M_g/L_g$ ratio is under control,
Fig.~\ref{f9} shows a discriminatory prediction which can be compared
with data. Once again, the problem concerns both the evolution of the
mean $M_g/L_g$ ratio and single galaxy deviations from average.
Supposing that the average $M_g/L_g$ evolution is under control, we
can estimate the impact of individual deviations by replacing the
sharp threshold on $M_g$ with a soft threshold, substituting each
galaxy mass $M_g$ with $M_g + \Delta M_g\, R$, as in the previous
section.  This test was performed for samples as wide as those in our
100$\, h^{-1}$Mpc box and the effects of such replacement are modest,
amounting to $\sim 10\, \%$ of the difference found between
$\Lambda$CDM and SUGRA.
 
Accordingly, the critical issue concerns the mean $M_g/L_g$ ratio.  By 
comparing $VPF$'s for different thresholds, we can however see that, 
in order that the shift of $M_{tr}$ induces a VPF shift similar to 
differences between models, $M_{tr}$ is to be displaced by a factor 
1.7--1.8. Uncertainties $\sim 10\, \%$ on the mean $M_g/L_g$ 
ratio would therefore leave intact the discriminatory power of the VPF 
statistics. 
 
Before concluding this section, let us finally comment on sample 
variance. All $\Lambda$CDM VPF's were estimated on the $\Lambda$CDM1 
simulation.  Differences between $\Lambda$CDM1 and $\Lambda$CDM2 are 
however small and could not be appreciated in the above 
plots. Accordingly, sample variance is not a relevant limit to the use 
of VPF's.

\begin{figure} 
\centering 
\epsfxsize=\hsize\epsffile{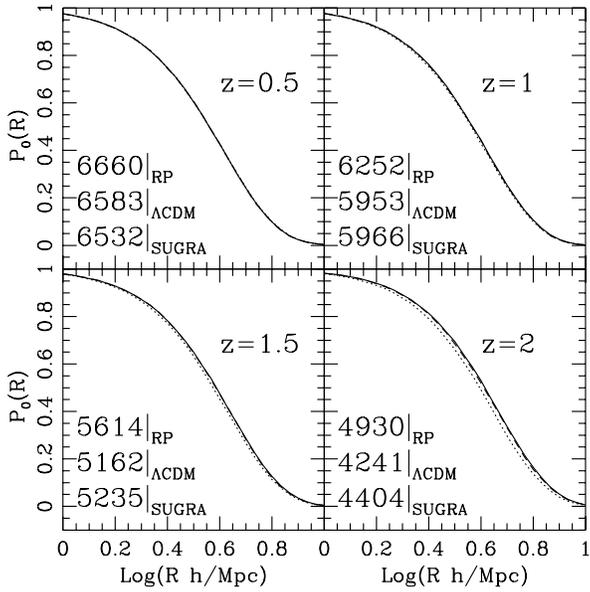} 
\caption{ VPF's in comoving volumes for $M_{tr} = 6 \cdot 10^{11} 
M_\odot h^{-1}$.  The four panels refer to redshift 0.5, 1, 1.5, 2, as 
indicated in the frames. The galaxy numbers in the simulation box are 
reported, for each model and redshift. Solid, dashed and dotted lines 
as in Fig.~\ref{f6}. 
} 
\label{f6ter}   
\end{figure}   
\begin{figure} 
\centering 
\epsfxsize=\hsize\epsffile{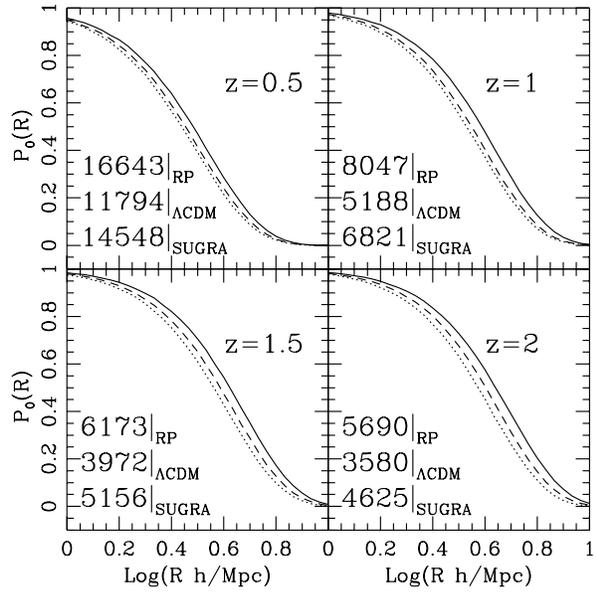} 
\caption{ VPF's in comoving volumes for fixed angular density.  The 
four panels refer to redshift 0.5, 1, 1.5, 2, as indicated in the 
frames. Numbers in the frames as in Fig.~\ref{f6ter}. Solid, dashed 
and dotted lines as in Fig.~\ref{f6}.} 
\label{f7}   
\end{figure}   
\begin{figure} 
\centering 
\epsfxsize=\hsize\epsffile{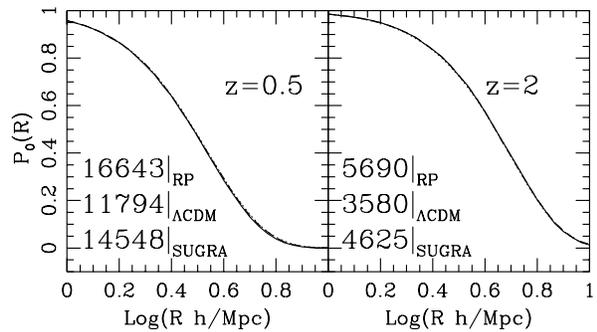} 
\caption 
{VPF's in fiducial volumes for fixed angular density.   
Numbers in the frames as for Fig.~\ref{f6ter}. Solid, dashed and 
dotted lines as in Fig.~\ref{f6}.  } 
\label{f8d} 
\end{figure} 
\nobreak 
\begin{figure} 
\centering 
\epsfxsize=\hsize\epsffile{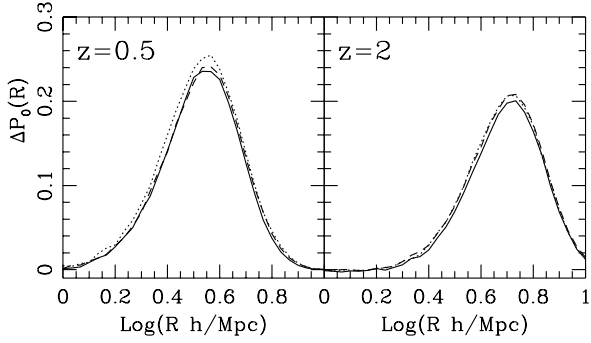} 
\caption 
{Differences between VPF's of various models and the VPF for a Poisson 
sample, in fiducial volumes for fixed angular density. They arise from 
the sum of $n$--point correlation functions in eq.~(\ref{po}). Tiny 
residual differences between models, almost unappreciable in the 
previous plot, arise from differences between their $n$--point 
functions, clearly almost erasen by geometrical renormalization. 
 } 
\vskip -.7truecm 
\label{f8v} 
\end{figure} 
\begin{figure} 
\centering 
\epsfxsize=\hsize\epsffile{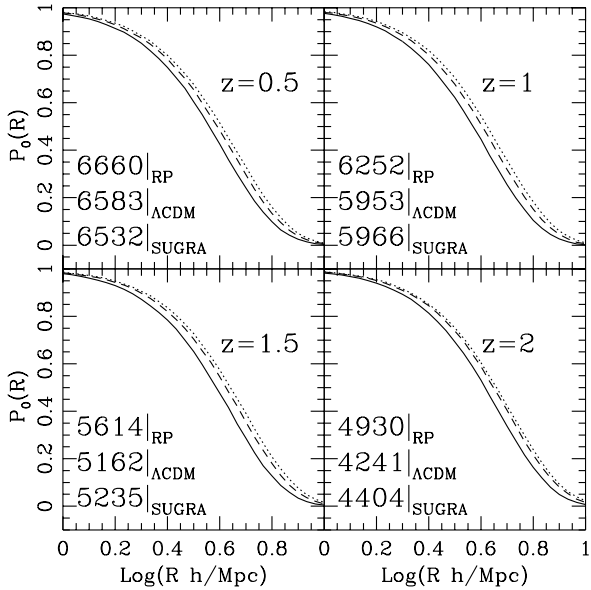} 
\caption{ 
VPF's in fiducial volumes for fixed mass limit (see text).   
Numbers in the frames as for Fig.~\ref{f6ter}. Solid, dashed and dotted lines 
as in Fig.~\ref{f6}.}   
\label{f9}  
\vskip -.1truecm  
\end{figure}   
   
\section{Conclusions}   

Dark Energy modifies the rate of cosmic expansion in the epoch when a   
substantial fraction of fluctuations on cluster scales reach their   
turnaround. Therefore, it seems quite natural to trace the redshift   
dependence of $w(z)=p_{de}/\rho_{de}$ using the cluster mass function   
at different redshifts.   

Unfortunately, the situation is more complicated.  The evolution of
$w(z)$ affects observations in two ways. First, it causes objects to
form and evolve with different rate. Second, it results in a different
mapping of comoving coordinates of galaxies to observed angular
positions and redshifts. Then, on a cluster mass scale ($\sim 10^{14}
h^{-1} M_\odot$), the evolutionary and geometrical effects tend to
cancel, which makes clusters somewhat problematic for testing the
equation of state.
   
In this paper we discuss the use of scales where the evolutionary
effects are minimized, so that the geometrical effects leave a clearer
imprint.  In a sense, this is not a new procedure: an analogous idea
is utilized when the DE equation of state is tested by using a {\it
standard candle}.  Obviously, galaxies are not standard candles
themselves, but they can provide a ``standard meter'' through their
(almost) model--independent abundance and evolution.
   
Previous analysis, which focused on clusters, tried to overcome the
above difficulties by making recourse to various features.  For
example, the evolutionary dependence on $w$ is preserved if one
considers masses {\it well} above $10^{14} h^{-1} M_\odot$.
Unfortunately, clusters with masses $\sim 10^{15} h^{-1} M_\odot$ or
larger are rare today and surely are even more rare in the past. Some
analysis (see, {\it e.g.}, Haiman et al. 2000) stressed a possible
role of very massive clusters at large $z$. Yet, the number of such
clusters cannot be large and, thus, comparing such predictions with
observations is a hard challenge. There is also another problem with
using cluster masses.  Usually the mass function is estimated with a
PS--like approximations, which are well tested with simulations. The
``virial'' radius $R_v$ is defined so that inside $R_v$ the density
contrast is $\Delta_v$. The value of $\Delta_v$ depends on the
redshift and on DE model. On the contrary, data are typically analized
with a standard density contrast $\Delta_c \simeq 180$ (or
200). Increasing $\Delta_c$ reduces the amplitude of the mass
function. If mass functions defined with variable $\Delta_v$ (almost)
overlap one another, mass functions defined with constant $\Delta_c$
can be different. In order to account for these differences in the
definitions, one needs to assume some shape for the density profile in
the outskirts of clusters. This is typically done by using a NFW
profile with concentration $c_s \simeq 5$. As we deal with rather
peripheral (virial) cluster regions, we can neglect the spread of
actual values of concentration.  However, when different $w(z)$ are
considered, the {\it mean} $c_s$ changes substantially, up to 80$\,
\%$ (Klypin et al 2003; Kuhlen et al. 2005). The differences between
$M_{200}$ and $M_{\rm v}$ are not large -- 10--$15\,
\%$. Approximately the same percent of the difference depends on
$w$. This may be still important. Neglecting these corrections may
lead to substantial systematic errors.

As an alternative to using galaxy clusters, so avoiding these and
other problems, here we suggest to exploit the dependence on DE nature
of the redshift distribution of galaxies and argue that the
difficulties of this approach can be overcome.

A first problem is that
one needs to know how to treat subhalos of more massive halos because
a large fraction of galaxies is hosted by subhalos. To populate
massive halos with galaxies we use recent results on the halo
occupation distribution (HOD). Accordingly, we believe that
this difficulty can be readily overcome.

Dealing with galaxies also requires knowledge of their masses
$M_g$. The $M_g-L_g$ relation is more complex than the relation
between $M_v$, X-ray flux and $T$ in clusters.

Galactic evolution studies and/or techniques aiming to compare
dynamical or lensing masses with luminosities can be used to this aim
(Bressan et al 1994, Portinari et al 2004, Prada et al 2003).  It is
also known that there is no one--to--one correspondence between
$L_g$ and $M_g$, as the luminosities of two galaxies of the same mass
can be quite different, up to one order of magnitude.  We then
considered separately two different issues: (i) How well we must know
the evolution of $ \langle M_g / L_g \rangle_{L_g} $. (ii) Which can
be the impact of fluctuations about such average value, taking into
account that the distribution about average can depend on $z$ and
mass.

How precisely $\langle M_g/L_g\rangle_{L_g}$ is to be known, in order
that different cosmologies can be safely discriminated, is shown by
Fig.~\ref{f6}, according to Sec.~6: If we double the values of $\delta
\theta_{gg} /\theta_{gg}$ provided there, we have a fair estimate the
evolution rate ${\cal R} = \langle M_g/L_g \rangle_{L_g}(z)/ \langle
M_g/L_g \rangle_{L_g} (z=0)$ just covering the differences between
models. Typically, if the estimated $\langle M_g/L_g\rangle_{L_g}$
evolution is reliable at a $\sim 10\, \%$ level, different models
could be discriminated.

To reach this goal, fresh observational material is needed, but no
conceptual difficulty is apparently involved in its acquisition.

The spread of the $M_g/L_g$ ratio, around its average value, also
involves a delicate issue, as it could cause systematics. Here we
reported the effect of a random spread of $L_g$ in a luminosity
interval $L_1,L_2$ with $L_2 \simeq 20\, L_1$. If the physical
distribution of luminosities, for any mass and at any redshift, is
within these limits, then possible systematics are well within
errors due to other effects.

Further tests on $L_g$ spread were also performed, which are not
reported in detail in this paper. They apparently indicate that really
wide and {\it ad--hoc} distributions are necessary, in order that
possible systematics exceed Poisson uncertainty.

Although bearing these reserves in mind, we conclude that
estimates of the redshift dependence of the average $M_g/L_g$,
reliable within $\sim 10\, \%$, can enable us to obtain a fair
information on DE nature.

In this paper we also discuss various tests based on the void 
probability function. We find that the VPF is almost model independent 
when estimated for samples with constant angular number density of 
galaxies. This result apparently suggests that $n$--point functions 
are almost model independent, at any $z$, once distances are suitably 
rescaled. If the mean $M_g/L_g$ evolution is under control, at the 
10$\, \%$ level, we can also assess that VPF, for samples defined with 
a given $M_{tr}$, puts in evidence geometrical differences between 
models and provides a discriminatory statistics.

Combining the simulated halo distribution with the HOD provides an
effective tool for testing the equation of state of DE. More work is
needed to define the $L_g$--$M_g$ relation and, possibly, to reduce
systematic effects. It is justified to expect that the $z$--dependence
of galaxy distribution, in deep galaxy samples, will allow us to
constrain the DE nature even more reliably than the density of galaxy
clusters in future complilations, sampling them up to large $z$'s.

\vskip .3truecm   
\noindent   
{\bf Acknowledgments}: We thank Loris Colombo and Cesare Chiosi for
 stimulating discussions. An anonimous referee is also to be thanked
 for improving the arguments now in Sec.~6.  A.K. acknowledges
 financial support of NSF and NASA grants to NMSU. S.G.  acknowledges
 financial support by DAAD.  Our simulations were done at the LRZ
 computer center in Munich, Germany.

\appendix    
\section{Dynamical Dark Energy models}   
   
Dynamical DE is to be ascribed to a scalar field, $\phi$,   
self--interacting through an effective potential $V(\phi)$, whose   
dynamics is set by the Lagrangian density:   
\begin{equation}   
\label{eq:lag}   
{\cal L}_{DE}  = -{1\over2} \, \sqrt{-g} \,    
\left( \partial^\mu \phi \partial_\mu \phi + V(\phi) \right)~.   
\end{equation}   
Here $g$ is the determinant of the metric tensor $g_{\mu  \nu}=    
a^2(\tau) dx_{\mu} dx_{\nu}$ ($\tau $ is the conformal   
time). In this work we need to consider just a spatially homogeneous   
$\phi$ ($\partial_i \phi \ll   
\dot \phi$; $i = 1,2,3$; dots denote differentiation with respect   
to $\tau$); the equation of motion is then:   
\begin{equation}   
\ddot \phi +2{\dot a \over a} \dot \phi+ a^2 {dV \over {d \phi}} = 0~.   
\label{eq:motion}    
\end{equation}   
Energy density and pressure, obtained from the   
energy--momentum tensor $T_{\mu \nu}$, are:   
\begin{equation}   
\label{eq:prho}   
\rho = -T^0_0 =  {\dot \phi^2\over {2a}} + V(\phi)~,   
~~~~~   
p ={1\over 3} \, T^i_i = {\dot \phi^2\over {2a}} - V(\phi)~,   
\end{equation}   
so that the state parameter   
\begin{equation}   
\label{eq:wde}   
w \equiv {p \over \rho} ={   
{ {\dot\phi^2/{2a}} - V(\phi)}\over { {\dot\phi^2/{2a}}    
+ V(\phi)}}   
\end{equation}   
changes with time and is negative as soon as the potential term $V(\phi)$ takes   
large enough values.   
   
The evolution of dynamical DE depends on details of the effective   
potential $V(\phi)$. Here we use the model proposed by Ratra \&   
Peebles (1988), that yield a rather slow evolution of $w$, and the   
model based on supergravity (Brax \& Martin 1999, 2001;   
 Brax, Martin \& Riazuelo 2000). The latter gives  a much faster   
evolving $w$.  The RP and SUGRA potentials    
\begin{equation}   
V(\phi) = \frac{\Lambda^{4+\alpha}} {\phi^\alpha} \qquad RP,   
\label{eq:1a}	   
\end{equation}   
\begin{equation}   
V(\phi) = \frac{\Lambda^{4+\alpha}}{\phi^\alpha} \exp (4\pi G   
\phi^2)~~~ SUGRA   
\label{eq:2a}	   
\end{equation}   
 cover a large spectrum of evolving $w$.     
These potentials allow tracker solutions, yielding the same low--$z$   
behavior that is almost independent of initial conditions.   
In eqs.~(\ref{eq:1a}) and (\ref{eq:2a}),    
$\Lambda$ is an energy scale  in the range   
$10^2$--$10^{10}\, $GeV, relevant for the physics of fundamental   
interactions. The potentials depend also on the exponent $\alpha$.   
Fixing $\Lambda$ and $\alpha$, the DE density parameter $\Omega_{de,o}$   
is determined. Here we rather use $\Lambda$ and $\Omega_{de,o}$ as   
independent parameters. In particular, numerical results are given   
for $\Lambda=10^3$GeV.   
   
The RP model with such $\Lambda$ value is in slight disagreement with low-$l$   
multipoles of the CMB anisotropy spectrum data. Agreement may be recovered   
with smaller $\Lambda$'s, which however loose significance in particle   
physics. The SUGRA model considered here, on the contrary, is in fair   
agreement with all available data.

   
\end{document}